\documentclass{PoS}

\title{Fluctuation and low transverse momentum correlation results from PHENIX}

\ShortTitle{Fluctuations and low transverse momentum correlation results from PHENIX}

\author{\speaker{Jeffery Mitchell}%
         \thanks{for the PHENIX Collaboration.}\\
        BNL, USA\\
        E-mail: \email{mitchell@bnl.gov}}

\abstract{The PHENIX Experiment at the Relativistic Heavy Ion Collider has conducted a survey of fluctuations in charged hadron multiplicity in Au+Au and Cu+Cu collisions at $\sqrt{s_{NN}}$ = 22, 62, and 200 GeV. A universal power law scaling for multiplicity fluctuations expressed as $\sigma^2/\mu^2$ is observed as a function of $N_{part}$ for all species studied that is independent of the transverse momentum range of the measurement.  PHENIX has also measured transverse momentum correlation amplitudes in p+p, d+Au, and Au+Au collisions. At low transverse momentum, significant differences in the correlations between the baseline p+p and d+Au data and the Au+Au data are presented.}

\FullConference{Correlations and Fluctuations in Relativistic Nuclear Collisions\\
                 July 7-9 2006\\
                 Florence, Italy}

\begin{document}

\section{A Survey of Charged Hadron Multiplicity Fluctuations}

The topic of event-by-event fluctuations of the inclusive charged particle multiplicity in relativistic heavy ion collisions has been revived by the observation of non-monotonic behavior in the scaled variance as a function of system size at SPS energies \cite{na49MF}.  The scaled variance is defined as $\sigma^{2}/\mu$, where $\sigma^{2}$ represents the variance of the multiplicity distribution in a given centrality bin, and $\mu$ is the mean of the distribution. For reference, the scaled variance of a Poisson distribution is 1.0, independent of $\mu$. PHENIX has surveyed the behavior of inclusive charged particle multiplicity fluctuations as a function of centrality and transverse momentum in $\sqrt{s_{NN}}$ = 62 GeV and 200 GeV Au+Au and 22, 62, and 200 GeV Cu+Cu collisions in order to investigate if the non-monotonic behavior persists at RHIC energies.

Details about the PHENIX experimental configuration can be found elsewhere \cite{phenixNIM}. All of the measurements described here utilize the PHENIX central arm detectors. The maximum PHENIX acceptance of $|\eta|<0.35$ in pseudorapidity and $180^{o}$ in azimuthal angle is considered small for event-by-event measurements. However, the event-by-event multiplicities are high enough in RHIC heavy ion collisions that PHENIX has a competitive sensitivity for the detection of many fluctuation signals. For example, a detailed examination of the PHENIX sensitivity to temperature fluctuations derived from the measurement of event-by-event mean $p_{T}$ fluctuations is described in \cite{ppg005}.

\begin{figure}
\resizebox{0.5\textwidth}{!}{%
  \includegraphics{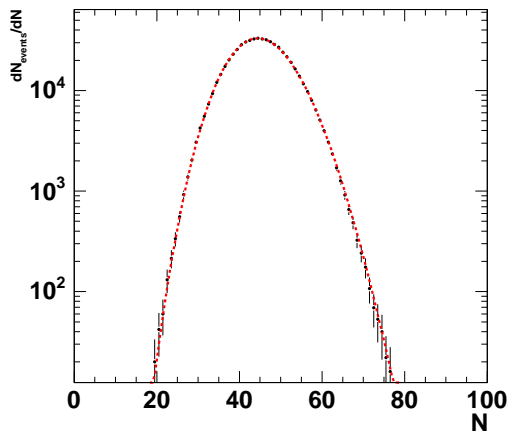}
}
\caption{PHENIX Preliminary. The event-by-event inclusive charged hadron multiplicity distribution for 0-5\% central 200 GeV Au+Au collisions with transverse momentum in the range $0.2<p_T<2.0$ GeV/c.  The dashed line is a Negative Binomial Distribution fit to the data.}
\label{fig:nbdFit}       
\end{figure}

It has been demonstrated that charged particle multiplicity fluctuation distributions in elementary and heavy ion collisions are well described by negative binomial distributions (NBD) \cite{e802MF}. The NBD of an integer $m$ is defined by
\begin{equation}
P(m) = \frac{(m+k-1)!}{m!(k-1)!} \frac{(\mu/k)^{m}}{(1+\mu/k)^{m+k}}
\end{equation}
where $P(m)$ is normalized for $0\leq m \leq \infty$, $\mu\equiv<m>$. The NBD contains an additional parameter, $k$, when compared to a Poisson distribution. The NBD becomes a Poisson distribution in the limit $k\rightarrow\infty$. The variance and the mean of the NBD is related to $k$ by $1/k = \sigma^{2}/\mu^{2} - 1/\mu$. The PHENIX multiplicity distributions are well described by NBD fits for all species, centralities, and transverse momentum ranges. The data presented here are results of NBD fits of the multiplicity distributions, an example of which is shown in Fig. \ref{fig:nbdFit}.

%

Each 5\% wide centrality bin selects a range of impact parameters. This introduces a component to the multiplicity fluctuations that can be attributed to fluctuations in the geometry of the collision. It is desireable to estimate and remove this known source of fluctuations so that only fluctuations due to the dynamics of the collision remain. The contribution of geometrical fluctuations is estimated using the HIJING event generator \cite{HIJING}, which well reproduces the mean multiplicity of RHIC collisions \cite{PPG019}.  The estimate is performed by comparing fluctuations from events with a fixed impact parameter to fluctuations from events with a range of impact parameters covering the width of each centrality bin, as determined from Glauber model simulations. The HIJING estimates are confirmed by comparing the HIJING fixed/ranged fluctuation ratios to measured 1\%/5\% bin width fluctuation ratios. The magnitude of the scaled variance is reduced by this correction. A 15\% systematic error for this estimate is included in the errors shown.

\begin{figure}[h]
\begin{minipage}{18pc}
\includegraphics[width=18pc]{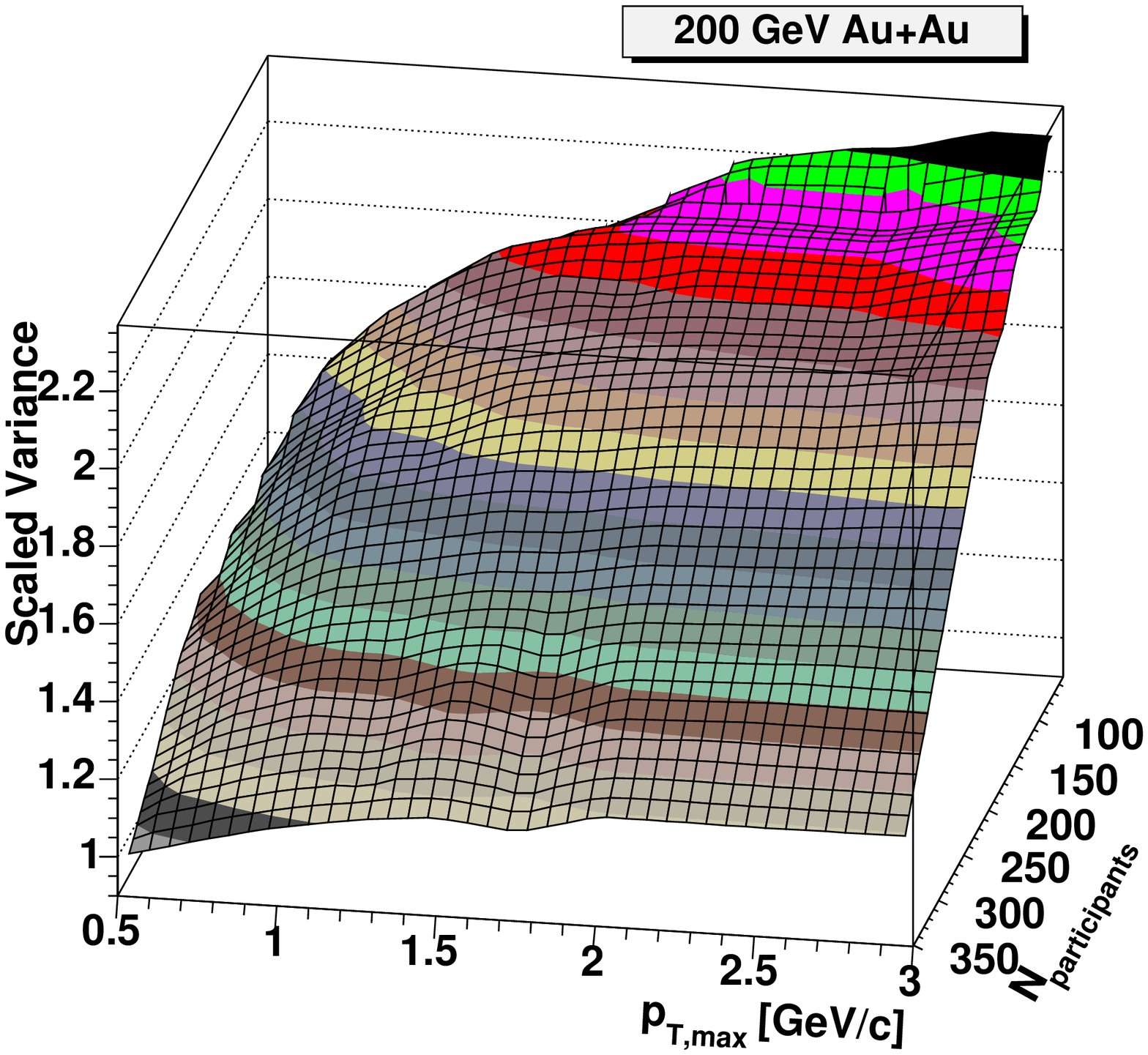}
\caption{\label{fig:auau200SvarGrid} PHENIX Preliminary multiplicity fluctuations for inclusive charged hadrons in terms of $\sigma^{2}/\mu$ as a function of $N_{participants}$ and the transverse momentum range $0.2<p_T<p_{T,max}$ GeV/c for 200 GeV Au+Au collisions.  The data have been corrected to remove impact parameter fluctuations and have been extrapolated to 2$\pi$ azimuthal acceptance.}
\end{minipage}\hspace{2pc}%
\begin{minipage}{18pc}
\includegraphics[width=18pc]{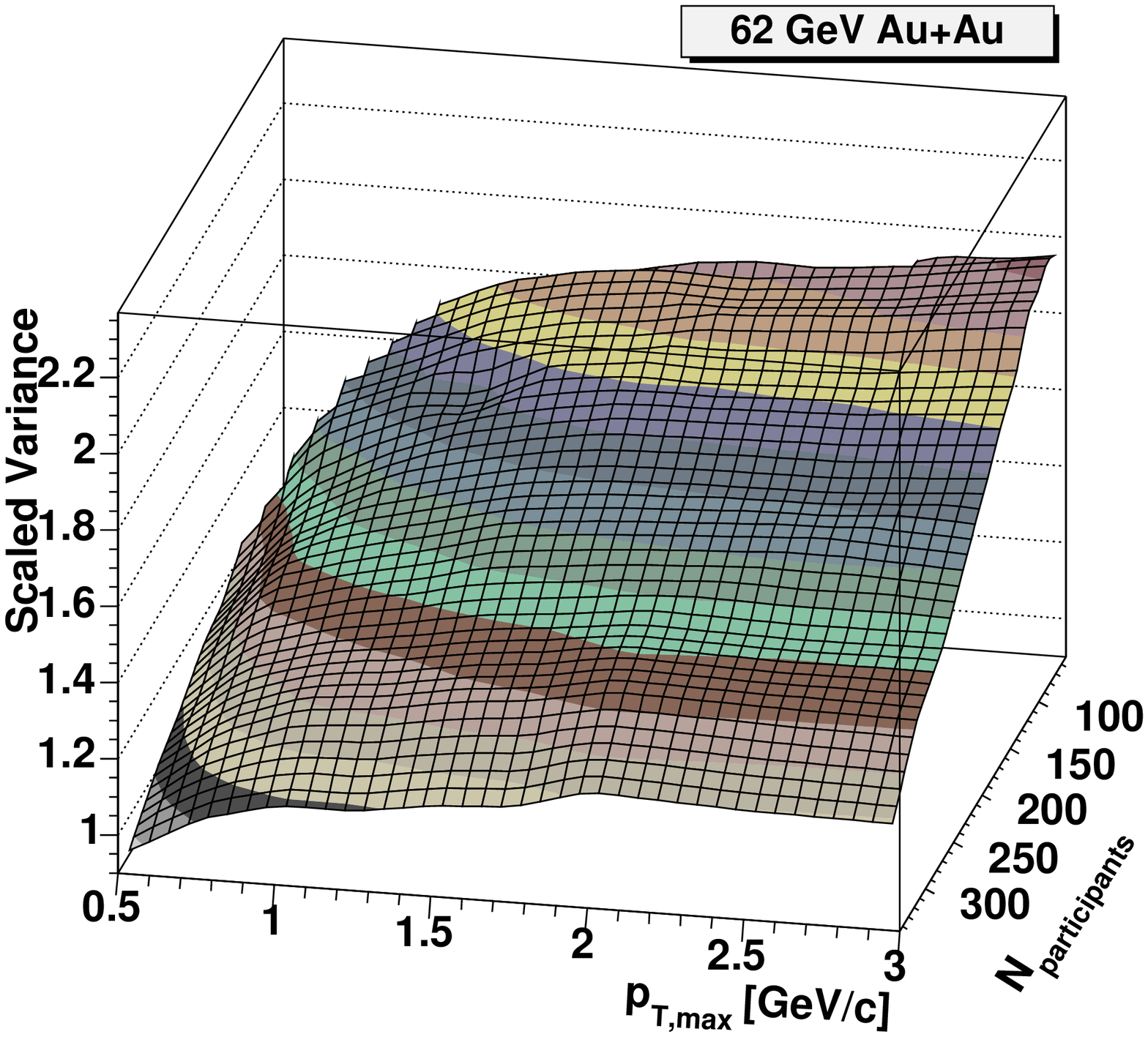}
\caption{\label{fig:auau062SvarGrid} PHENIX preliminary multiplicity fluctuations for inclusive charged hadrons in terms of $\sigma^{2}/\mu$ as a function of $N_{participants}$ and the transverse momentum range $0.2<p_T<p_{T,max}$ GeV/c for 62 GeV Au+Au collisions.  The data have been corrected to remove impact parameter fluctuations and have been extrapolated to 2$\pi$ azimuthal acceptance.}
\end{minipage}\hspace{2pc}%
\begin{minipage}{18pc}
\includegraphics[width=18pc]{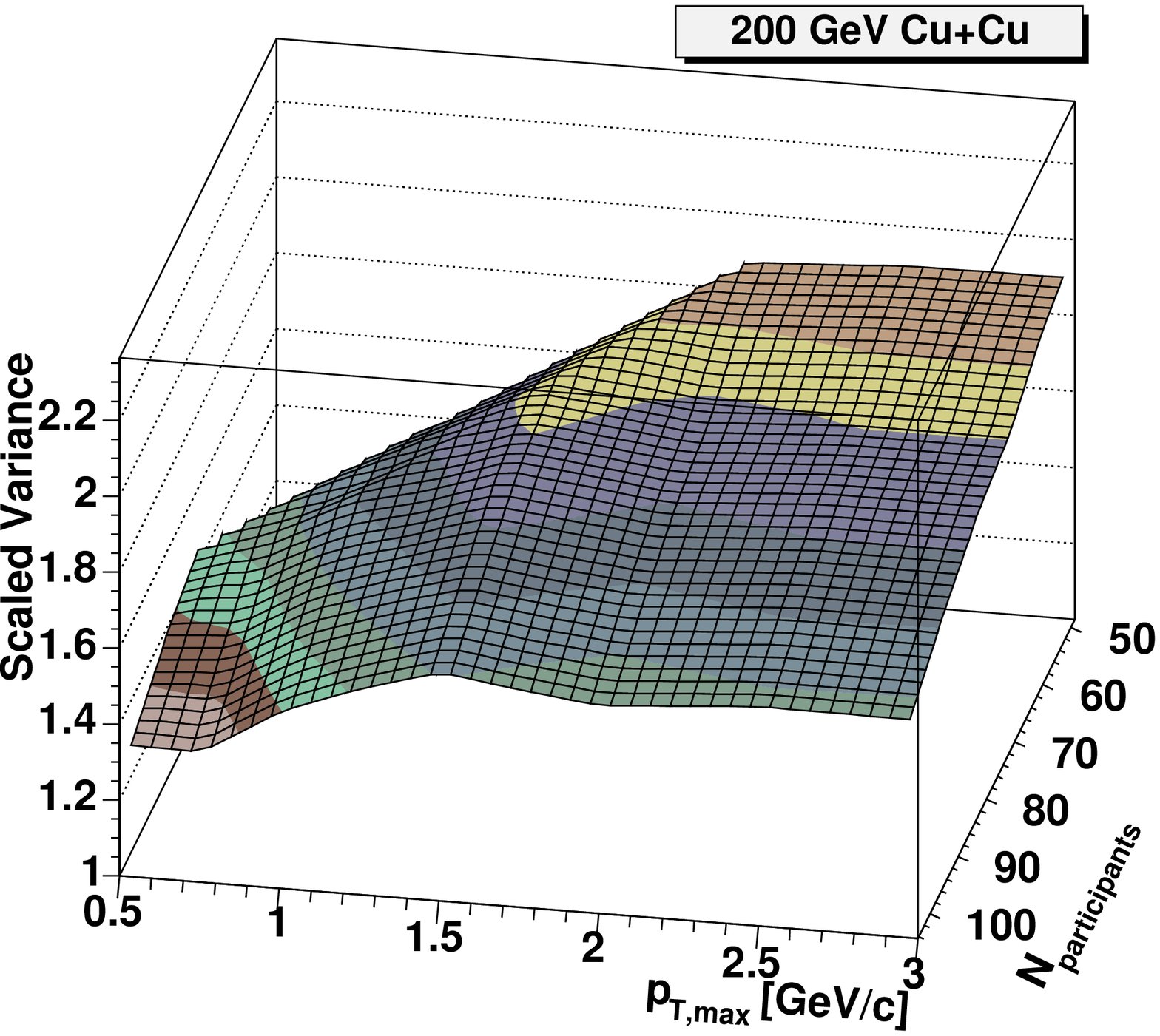}
\caption{\label{fig:cucu200SvarGrid} PHENIX preliminary multiplicity fluctuations for inclusive charged hadrons in terms of $\sigma^{2}/\mu$ as a function of $N_{participants}$ and the transverse momentum range $0.2<p_T<p_{T,max}$ GeV/c for 200 GeV Cu+Cu collisions.  The data have been corrected to remove impact parameter fluctuations and have been extrapolated to 2$\pi$ azimuthal acceptance.}
\end{minipage}\hspace{2pc}%
\begin{minipage}{18pc}
\includegraphics[width=18pc]{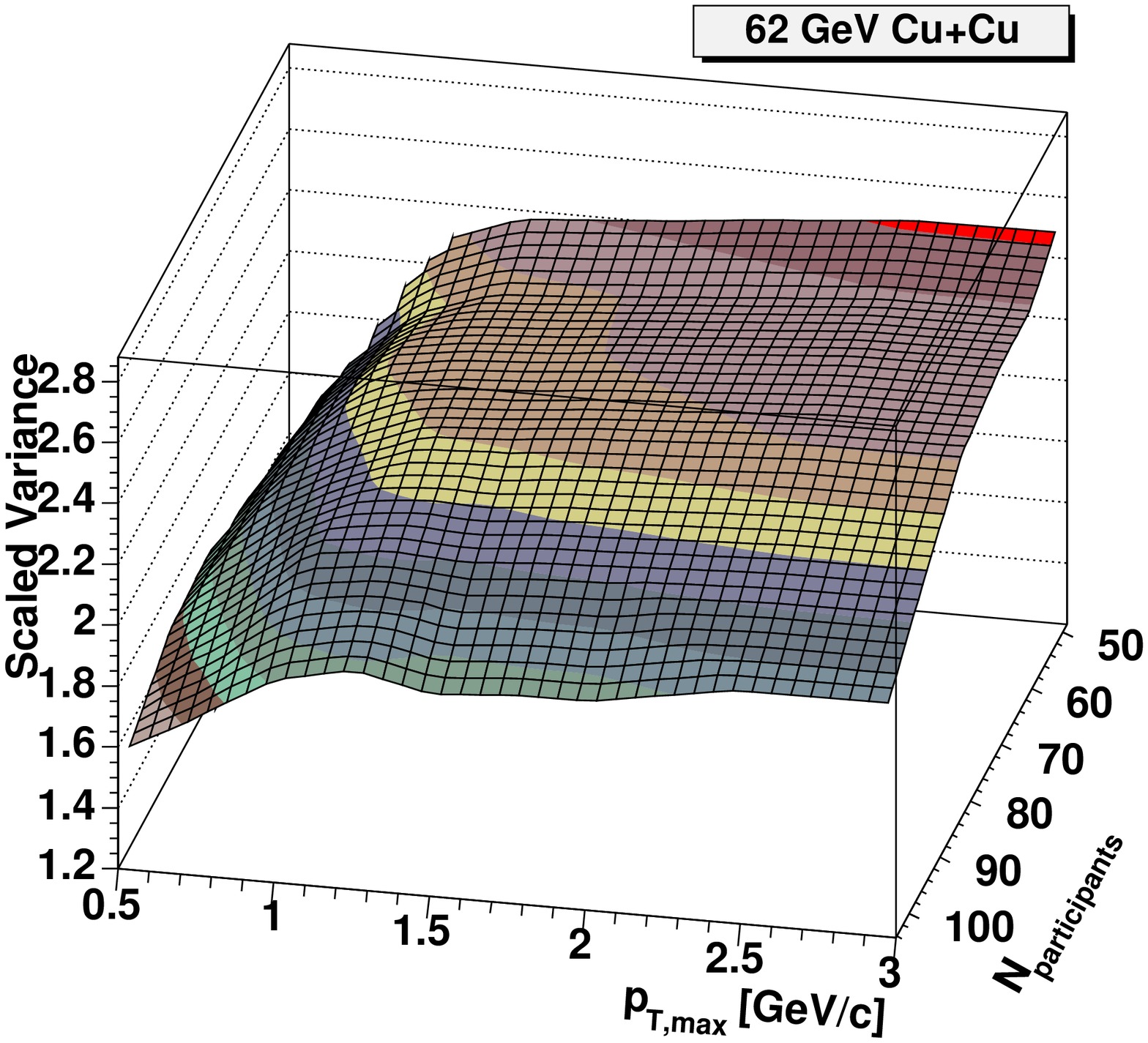}
\caption{\label{fig:cucu062SvarGrid} PHENIX preliminary multiplicity fluctuations for inclusive charged hadrons in terms of $\sigma^{2}/\mu$ as a function of $N_{participants}$ and the transverse momentum range $0.2<p_T<p_{T,max}$ GeV/c for 62 GeV Cu+Cu collisions.  The data have been corrected to remove impact parameter fluctuations and have been extrapolated to 2$\pi$ azimuthal acceptance.}
\end{minipage}\hspace{2pc}%
\end{figure}

By measuring the scaled variance in successively wider azimuthal ranges, a linear dependence on azimuthal acceptance is observed, with the scaled variance increasing with azimuthal acceptance.  In order to facilitate direct comparisons with other experiments, the multiplicity fluctuations quoted here have been linearly extrapolated to 2$\pi$ acceptance by fitting the azimuthal dependance within the detector acceptance. Systematic errors due to the extrapolation have been included in the total errors shown. The results of the survey are shown in Fig. \ref{fig:auau200SvarGrid} through \ref{fig:cucu062SvarGrid}, where the corrected scaled variance is plotted as a function of centrality and transverse momentum range.

\begin{figure}[h]
\begin{minipage}{18pc}
\includegraphics[width=18pc]{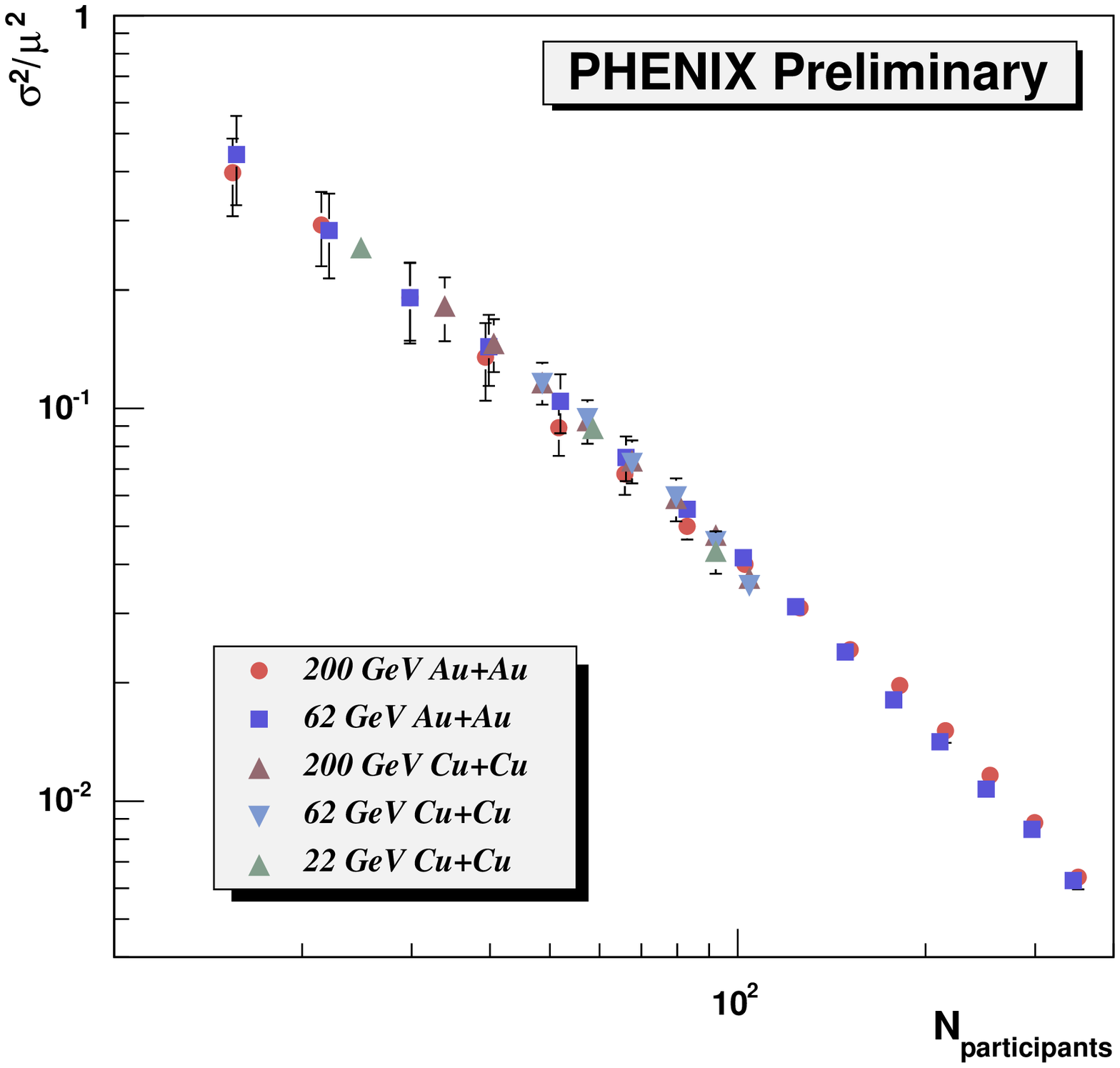}
\caption{\label{fig:s2v2VsCentPt7}Multiplicity fluctuations for inclusive charged hadrons in the transverse momentum range $0.2<p_T<2.0$ GeV/c in terms of $\sigma^{2}/\mu^{2}$ as a function of $N_{participants}$ for 200 GeV Au+Au, 62 GeV Au+Au, 200 GeV Cu+Cu, 62 GeV, and 22 GeV Cu+Cu collisions.  The data have been scaled by factors of 1.0, 0.73, 1.43, 0.73, and 1.0, respectively, in order to emphasize the universal scaling of all species. The dashed curve is a power law fit as described in the text.}
\end{minipage}\hspace{2pc}%
\begin{minipage}{18pc}
\includegraphics[width=18pc]{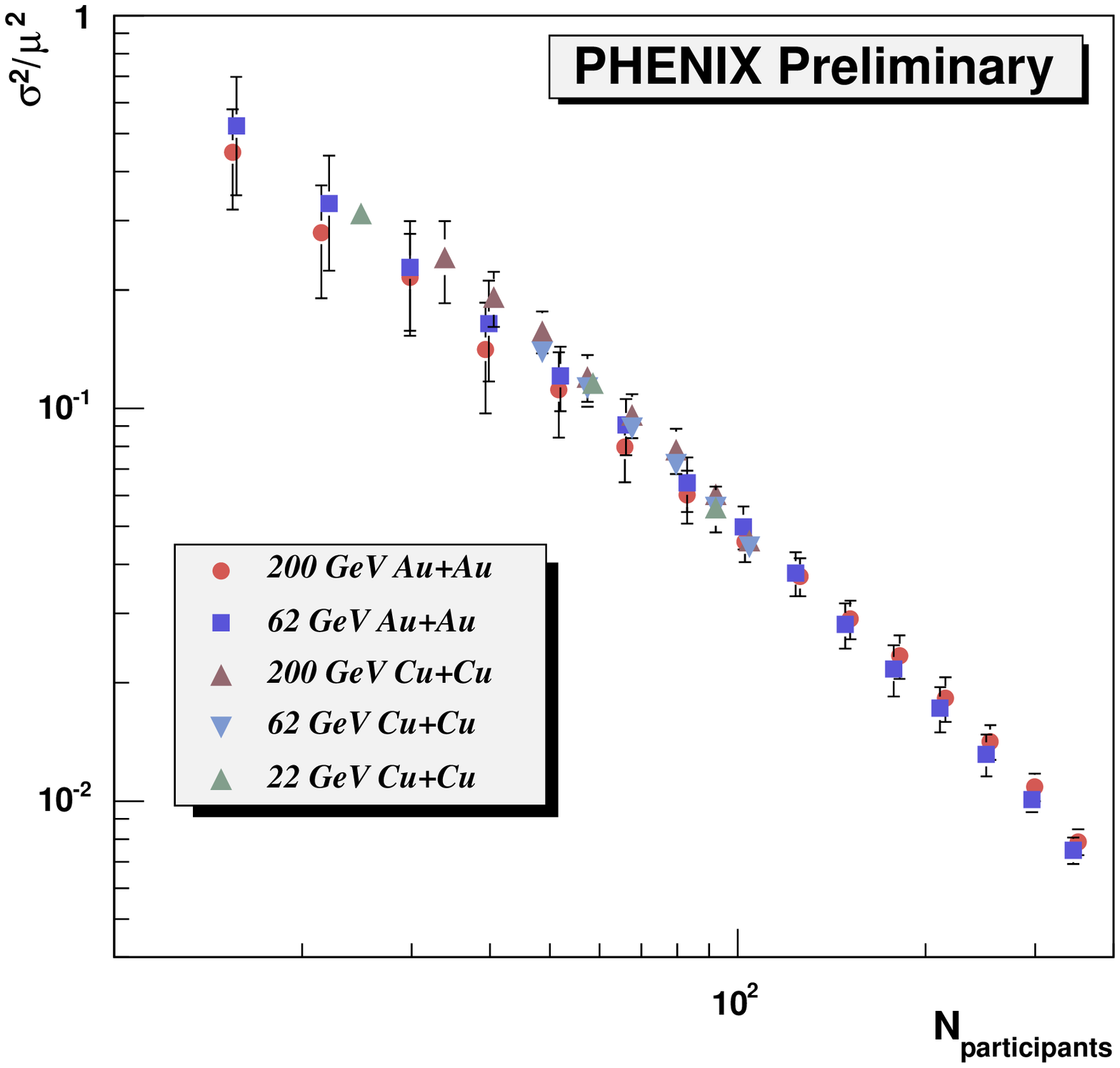}
\caption{\label{fig:s2v2VsCentPt2}Multiplicity fluctuations for inclusive charged hadrons in the transverse momentum range $0.2<p_T<0.75$ GeV/c in terms of $\sigma^{2}/\mu^{2}$ as a function of $N_{participants}$ for 200 GeV Au+Au, 62 GeV Au+Au, 200 GeV Cu+Cu, 62 GeV, and 22 GeV Cu+Cu collisions.  The data have been scaled by factors of 1.0, 0.73, 1.43, 0.73, and 1.0, respectively, in order to emphasize the universal scaling of all species. The dashed curve is a power law fit as described in the text.}
\end{minipage}\hspace{2pc}%
\end{figure}

In the Grand Canonical Ensemble, the variance and the mean of the particle number can be directly related to the compressibility: $\sigma^{2}/\mu^{2} = k_{B} (T/V) k_{T}$, where $k_{B}$ is Boltzmann's constant, T is the system temperature, and V is its volume \cite{Stanley}. The multiplicity fluctuations in terms of $\sigma^{2}/\mu^{2} = 1/k_{NBD}+1/\mu$ are shown in Fig. \ref{fig:s2v2VsCentPt7} as a function of centrality for Au+Au collisions at 200 and 62 GeV and Cu+Cu collisions at 200, 62, and 22 GeV.  In order to demonstrate their scaling properties as a function of centrality, each curve has been scaled to the 200 GeV Au+Au curve.  All four datasets can be described by a power-law function as follows: $\sigma^{2}/\mu^{2} \propto N_{part}^{-1.40 \pm 0.03}$. 

It is expected that the compressibility diverges as one approaches the critical point, and the rate of divergence is described by a power law, $k_{T} = A((T-T_{C})/T_{C})^{-\gamma}$, where $T_{C}$ is the value of the temperature at the critical point, and $\gamma$ is the critical exponent for isothermal compressibility \cite{Stanley}. However, it is not prudent to interpret the observed scaling as critical behavior since it has not been established that the temperature varies significantly as a function of $N_{part}$. A more robust critical exponent study could be performed with a RHIC scan at low energy where the freeze-out temperature could be estimated from identified particle spectra and Hanbury-Brown Twiss correlations, as outlined in \cite{phenixPhi}.

Figures \ref{fig:s2v2VsCentPt7} and \ref{fig:s2v2VsCentPt2} show the multiplicity fluctuations as a function of centrality for different $p_T$ ranges. In each case, the exponent of the power law function describing the data remains unchanged. Hence, the influence of $p_T$-dependent processes at higher $p_T$, such as hard scattering, have little effect on the scaling properties. The scaling also applies to the low energy 22 GeV Cu+Cu data.

\section{Low Transverse Momentum Correlations}

The technique of two-particle correlations has demonstrated its effectiveness in extracting information about jet production in reference (p+p and d+Au) and heavy-ion collisions at RHIC. Some features of the correlations are dependent on the $p_T$ of the trigger and associated particles.  It is important to examine and understand the evolution of correlation functions from high to low $p_T$ in heavy ion collisions with comparisons to reference collisions.  Examining the $p_T$ evolution of the correlations may provide information that can aid in the interpretation of such features such as excess jet production at low $p_T$ or eventually the observed displaced away side peaks.

Interpreting correlation functions at low $p_T$ (in this context, with both the trigger and associated particle below 1 GeV/c) is complicated by several factors: a) the hard scattering correlation becomes reduced in amplitude and diluted in azimuthal separation, b) correlations due to collective flow are present, although the magnitude of elliptic flow is much reduced at low $p_T$, c) Hanbury-Brown Twiss correlations are also present, and d) correlations due to resonance production are present. In order to help disentangle the latter two contributions, all correlations are separated into only like-sign or only unlike-sign pairs. Here, only the like-sign results, which should not have a significant resonance contribution, are presented.

The minimum bias datasets analyzed here contain 10 million events from the RHIC Run-3 p+p run, 10 million events from the Run-3 d+Au run, 2 million events from the Run-4 200 GeV Au+Au run, and 750,000 events from the Run-4 62 GeV Au+Au run.  All correlation functions are constructed using a strict mixed event procedure with all cuts, including track proximity cuts, applied identically to the data and mixed event samples. Mixed events are constructed from identical data classes (distinguished by centralities within 5\% and event vertex z coordinates within 10 cm) by directly sampling the data charged particle multiplicity distribution after all cuts have been applied.

Since every event contains multiple particles at low $p_T$, the correlations are constructed by considering every particle in the event as a trigger particle in succession and associating it with every other particle in the event. Correlations constructed in this manner are auto-correlations \cite{starAuto}. The correlation amplitudes reported here are normalized bin-by-bin as follows: $C = \frac{N_{data}/N_{events,data}}{N_{mixed}/N_{events,mixed}}$, where $N_{data}$ and $N_{mixed}$ are the number of pair counts in a given $p_T$ and/or $\Delta \phi$ bin. Correlations shown for the near-side are defined as the integrated correlation amplitude over the range $|\Delta\phi|<60^{o}$.

\begin{figure}[h]
\begin{minipage}{18pc}
\includegraphics[width=18pc]{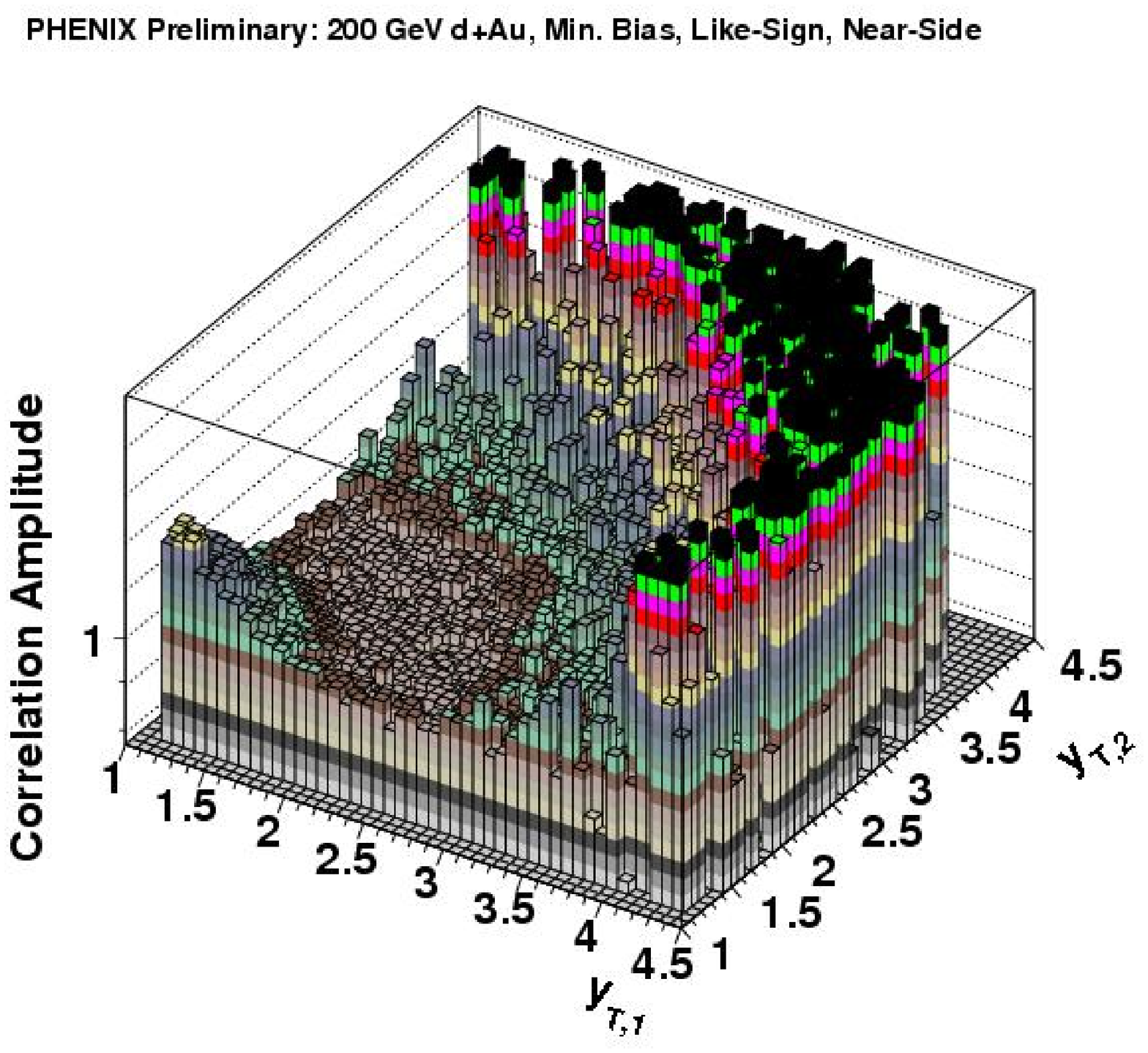}
\caption{\label{fig:dauytLSNear}Near-side correlation amplitudes for like-sign pairs as a function of the transverse rapidity of each particle in the pair for minimum bias 200 GeV d+Au collisions.}
\end{minipage}\hspace{2pc}%
\begin{minipage}{18pc}
\includegraphics[width=18pc]{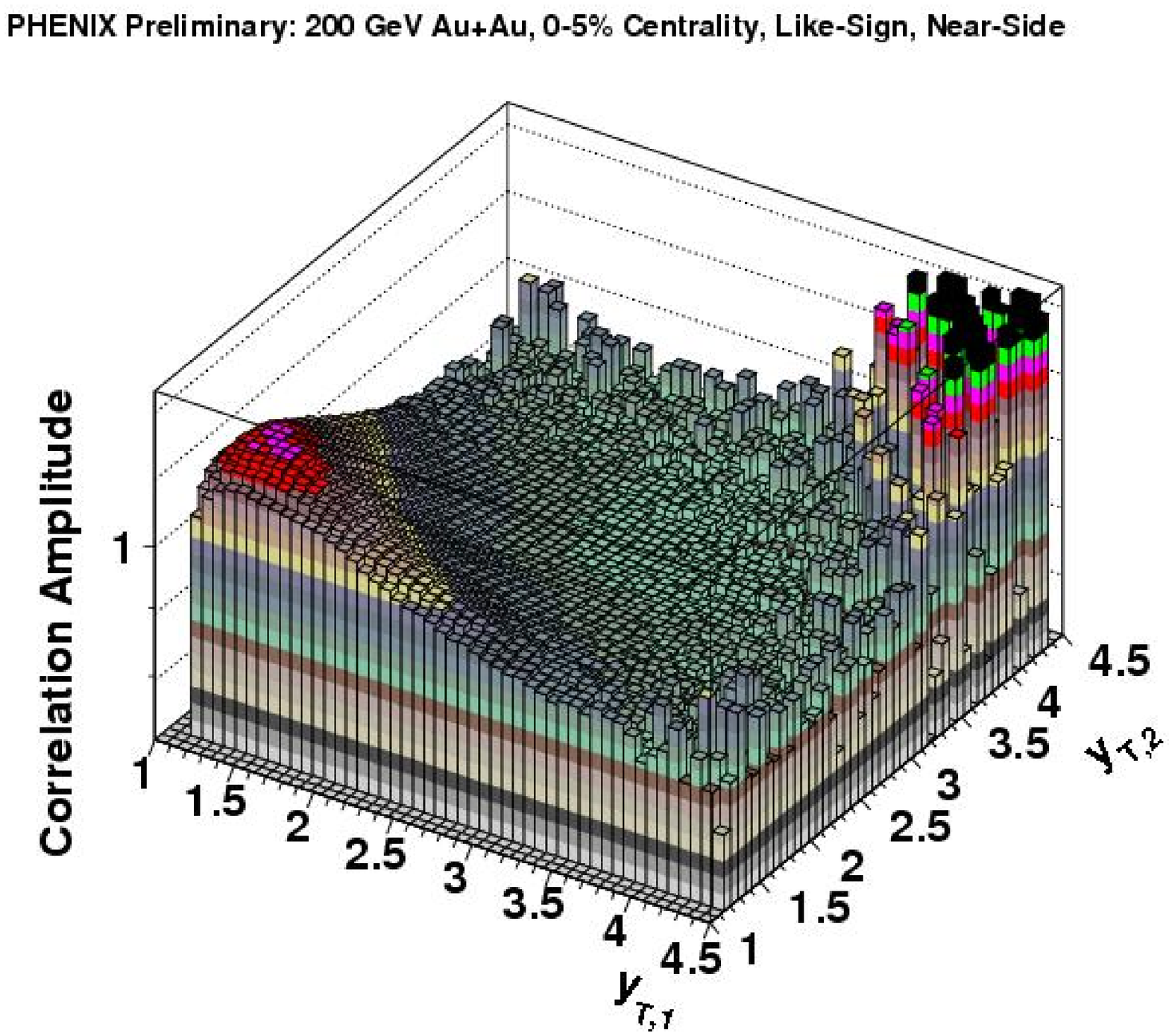}
\caption{\label{fig:auauytLSNear}Near-side correlation amplitudes for like-sign pairs as a function of the transverse rapidity of each particle in the pair for 0-5\% central 200 GeV Au+Au collisions.}
\end{minipage}\hspace{2pc}%
\end{figure}

\begin{figure}[h]
\begin{minipage}{18pc}
\includegraphics[width=18pc]{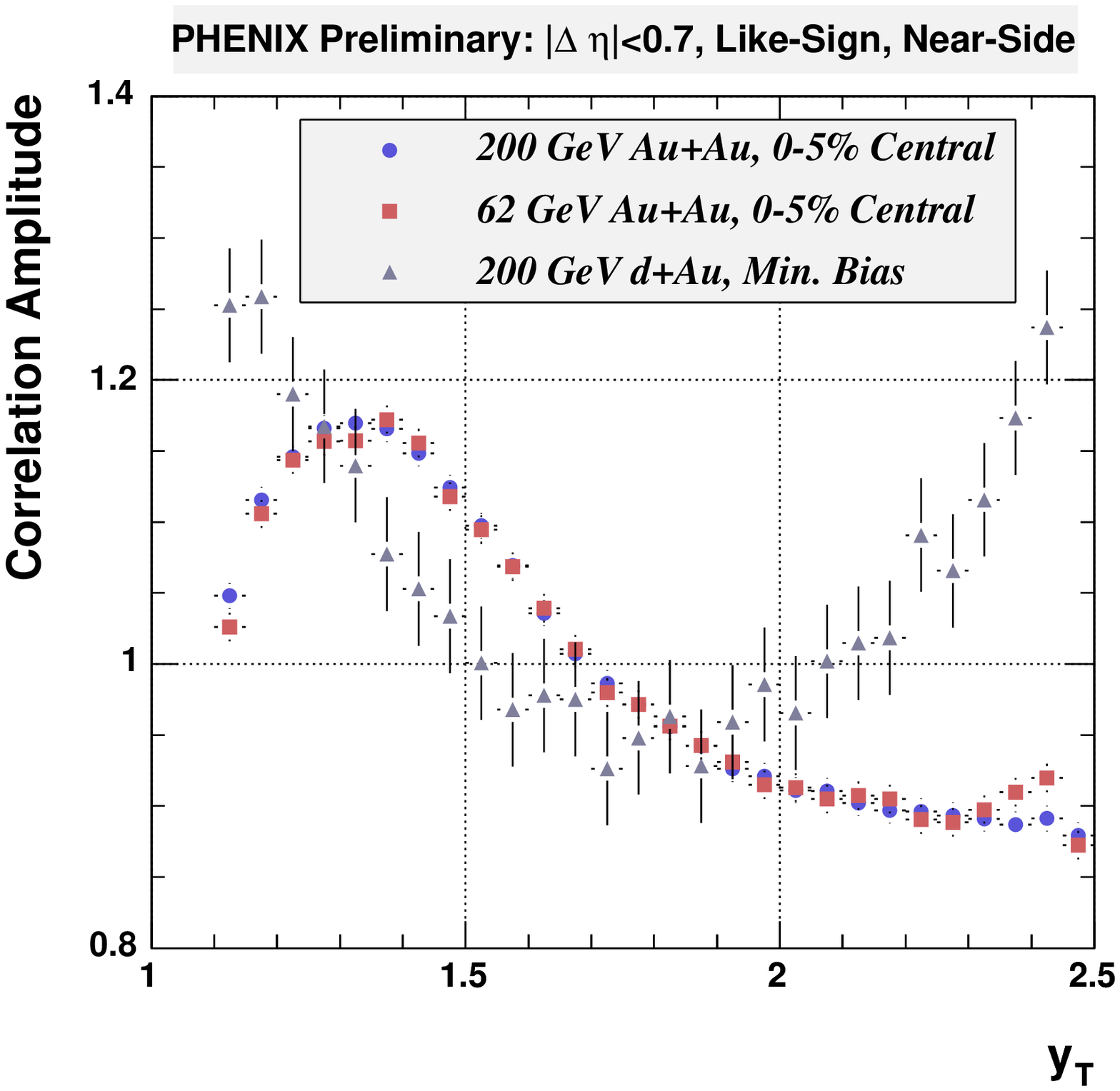}
\caption{\label{fig:ytDiagLSNS}Near-side correlation amplitudes for like-sign pairs with $p_T$ bins chosen such that $p_{T,1} = p_{T,2}$ for Au+Au and d+Au collisions. These correlation amplitudes include contributions from flow in the Au+Au collisions. The error bars include statistical and systematic errors.}
\end{minipage}\hspace{2pc}%
\begin{minipage}{18pc}
\includegraphics[width=18pc]{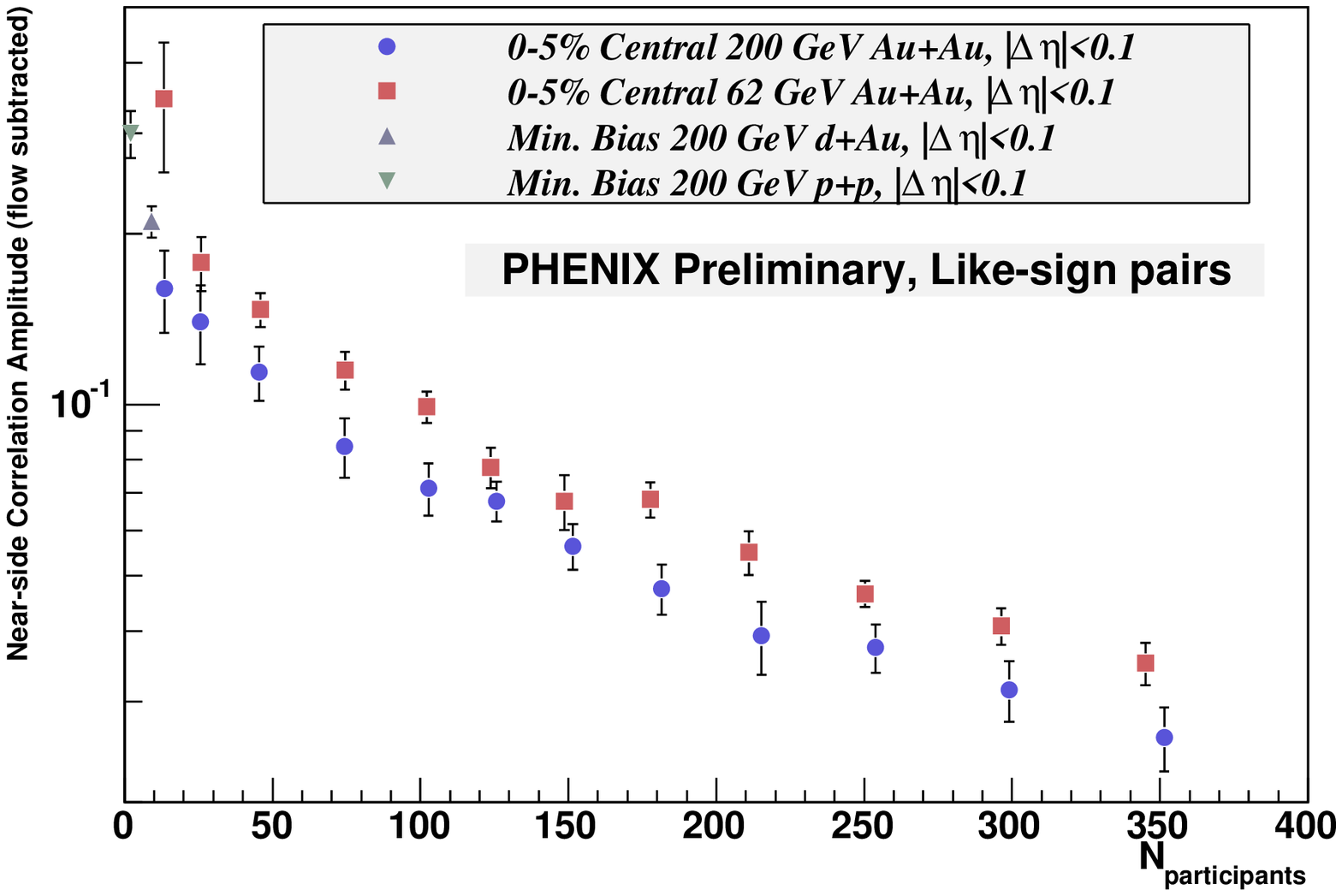}
\caption{\label{fig:ampVsCentLSNS}Near-side correlation amplitudes for like-sign pairs with $p_T$ between 200 and 500 MeV/c and $|\Delta \eta|<0.1$ for p+p, d+Au, and Au+Au collisions. Flow has been subtracted from the Au+Au data. The errors are dominated by systematic errors due to the fit procedure.}
\end{minipage}\hspace{2pc}%
\begin{minipage}{18pc}
\includegraphics[width=18pc]{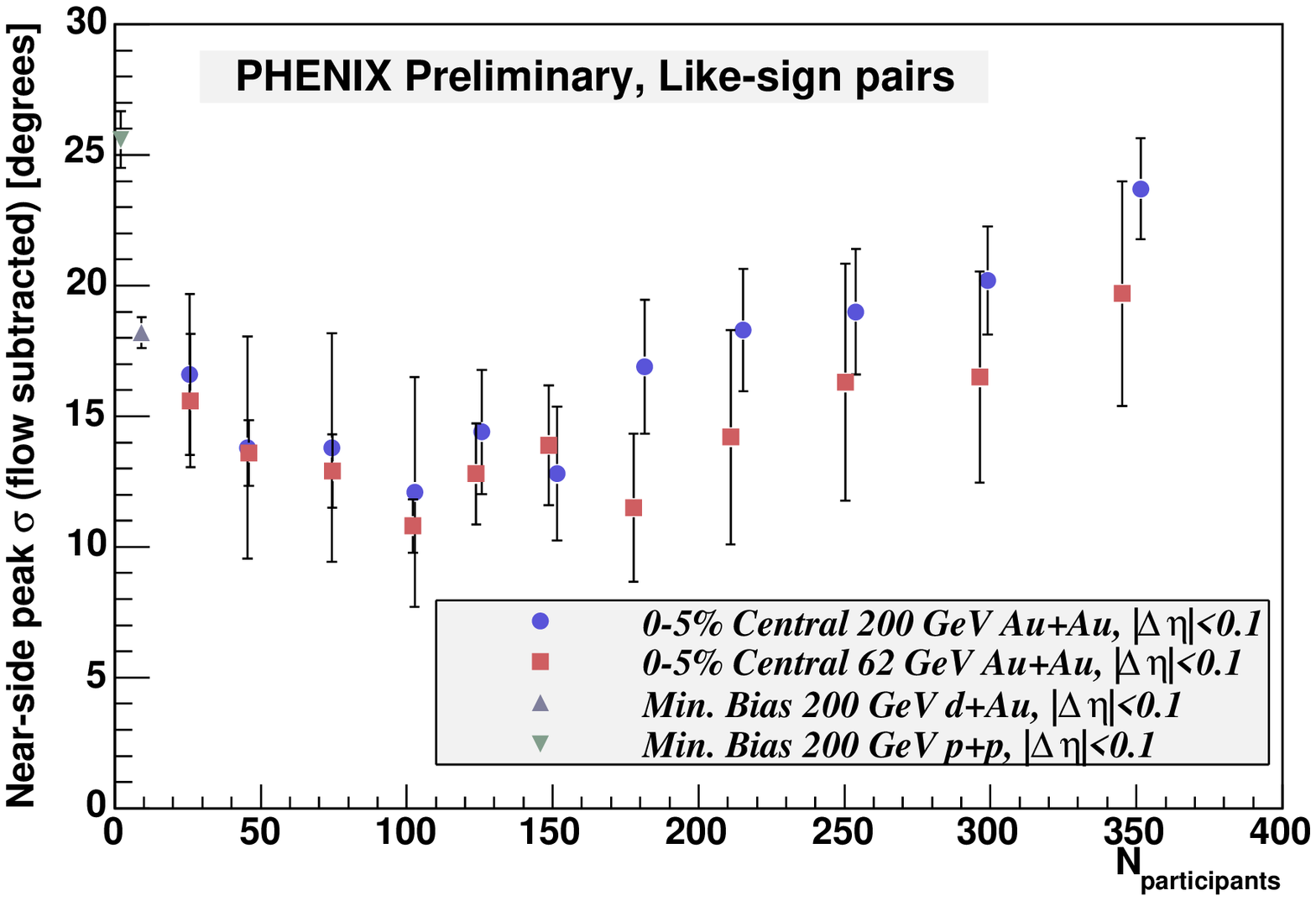}
\caption{\label{fig:sigmaVsCentLSNS}Near-side correlation widths for like-sign pairs with $p_T$ between 200 and 500 MeV/c and $|\Delta \eta|<0.1$ for p+p, d+Au, and Au+Au collisions. Flow has been subtracted from the Au+Au data. The errors are dominated by systematic errors due to the fit procedure.}
\end{minipage}\hspace{2pc}%
\begin{minipage}{18pc}
\includegraphics[width=18pc]{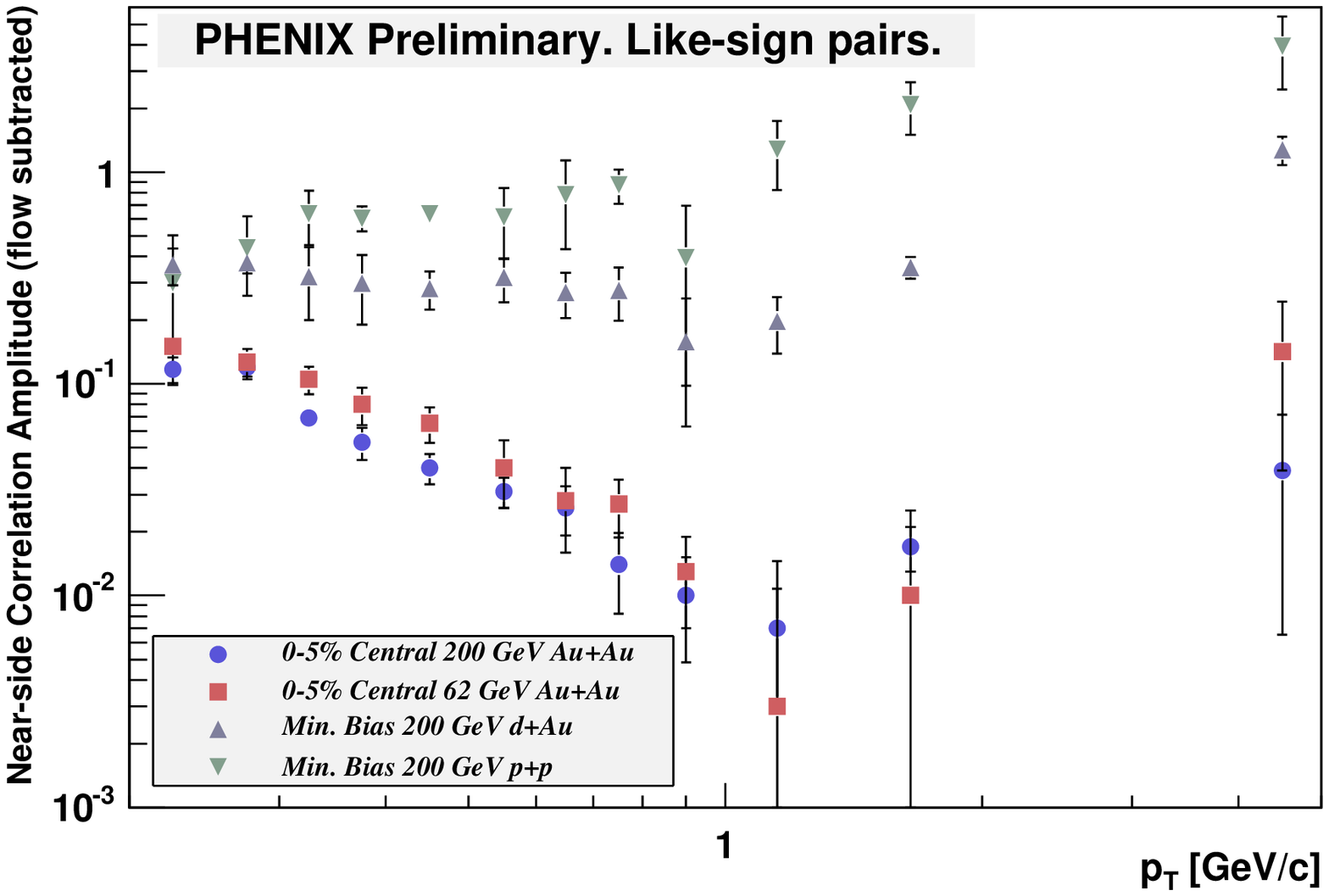}
\caption{\label{fig:ampVsPtLSNS}Near-side correlation amplitudes for like-sign pairs both with $p_T$ within the given $p_T$ bin and with $|\Delta \eta|<0.1$ for p+p, d+Au, and 0-5\% central Au+Au collisions. Flow has been subtracted from the Au+Au data. The errors are dominated by systematic errors due to the fit procedure.}
\end{minipage}\hspace{2pc}%
\end{figure}

Correlation amplitudes plotted as a function of the transverse rapidity, $y_T = ln((m_T + p_T)/m_0)$ with the pion mass assumed for $m_{0}$, for like-sign pairs on the near-side, are shown in Fig. \ref{fig:dauytLSNear} for minimum bias 200 GeV d+Au collisions and in Fig. \ref{fig:auauytLSNear} for 0-5\% central 200 GeV Au+Au collisions.  The $y_T$ variable is chosen in order to emphasize the distribution at low $p_T$ with respect to that at high $p_T$ and to facilitate comparisons to measurements by the STAR Collaboration \cite{starAuto}. For reference, $y_T=1.5$ corresponds to $p_T=300$ MeV/c and $y_T=2.7$ corresponds to $p_T=1.0$ GeV/c for pions. There are two primary features observed in these correlations.  First, there is the expected large correlation amplitude primarily due to hard scattering processes when either particle in the pair has a large $p_T$ along with a small fractional contribution due to elliptic flow. Second, there is a peak when both pairs have low $p_T$. This peak is dominated by HBT correlations, which has been confirmed by observing a sharp reduction in the amplitude of the peak for unlike-sign pairs and by simultaneously observing a peak in the $Q_{invariant}$ distribution in this region for like-sign pairs. Closer xamination of the low $p_T$ peak reveals that there is a difference in the location of the peak in $y_T$ between d+Au and Au+Au collisions. In d+Au collisions, the peak amplitude rises up to the low $p_T$ edge of the PHENIX acceptance. This holds for p+p collisions, also. However, there is a maximum in the peak at $y_T \approx 1.4$ ($p_T \approx 250$ MeV/c) in Au+Au collisions at both 200 and 62 GeV. Fig. \ref{fig:ytDiagLSNS} shows the peak taken along the $p_{T,1} = p_{T,2}$ diagonal for near-side like-sign pairs. Possible explanations for these differences require further study.

To facilitate a more detailed direct comparison of the d+Au and Au+Au correlations over the entire $p_T$ range, it is necessary to subtract the flow component from the Au+Au azimuthal correlation distributions. This is done by fitting the near-side region ($\Delta \phi < 90^{o}$) of these distributions with a function containing a harmonic $cos(2 \Delta \phi)$ term and a Gaussian distribution with a mean of $\Delta \phi = 0$. For p+p and d+Au collisions, the Gaussian fit is performed, but the harmonic term is not included. 

Fig. \ref{fig:ampVsCentLSNS} shows the correlation amplitude from the near-side Gaussian fit for like-sign pairs as a function of centrality. For this plot, all pairs have a $p_T$ between 200 and 500 MeV/c. In order to best isolate the HBT contribution to the peak, all pairs are required to be within $|\Delta \eta|<0.1$. For both 200 and 62 GeV Au+Au collisions, the amplitude drops exponentially as a function $N_{participants}$. Only a small fraction of the decrease in the amplitude is due to random dilution of the correlation amplitude \cite{vol04}.

Fig. \ref{fig:sigmaVsCentLSNS} shows the standard deviation of the near-side Gaussian fit for like-sign pairs as a function of centrality. Again, both particles in all pairs have a $p_T$ between 200 and 500 MeV/c and $|\Delta \eta|<0.1$. The error bars are dominated by systematic errors due to the fitting procedure. There is a large difference observed in the width between p+p and d+Au collisions. In Au+Au collisions, particularly for the higher statistics 200 GeV Au+Au dataset, a significant increase in the width is seen in the most central collisions. 

Fig. \ref{fig:ampVsPtLSNS} shows the correlation amplitude for the near-side Gaussian fit for like-sign pairs as a function of transverse momentum. Here, $p_{T,min}<p_{T,1}<p_{T,max}$ and $p_{T,min}<p_{T,2}<p_{T,max}$ ($p_T$ within the limits of the bin, $p_{T,min}$ and $p_{T,max}$) and $|\Delta \eta|<0.1$. The points are plotted in the geometric center of the bin. The bins are exponentially distributed in order to nullify any variations due to random dilution of the correlation amplitudes within each collision species. In both the p+p and d+Au datasets, the amplitudes are relatively flat as a function of $p_T$ up to about 1-1.5 GeV/c, where the amplitudes begin to rise significantly. This rise is likely due to the influence of correlations due to jets, which has been confirmed by the appearance of an away-side component in the azimuthal correlations. A rise in the correlation amplitudes for Au+Au collisions is also seen starting at the same $p_T$. However, unlike the p+p and d+Au datasets, the Au+Au data exhibit an exponential decrease in the amplitude as a function of $p_T$ below 1 GeV/c. This decrease is contrary to the trend in the reference datasets, and contrary to the ansatz of significant contributions from jet production at low $p_T$ \cite{starAuto}. More detailed studies of the cause of this effect are underway.

\section{Summary}

PHENIX has completed a comprehensive survey of charged hadron multiplicity fluctuations for several different collision species and energies, all of which demonstrate a universal power-law scaling as a function of $N_{part}$. Also presented is a survey of two-particle correlations for several collision species at 62 and 200 GeV focusing on the behavior of the near-side peak for like-sign pairs, which is dominated by HBT correlations. Examination of spatial HBT correlations shows several statistically significant differences between p+p and d+Au collisions and Au+Au collisions that will warrant further study.

\end{document}